\documentclass[%
 reprint,
superscriptaddress,
 amsmath,amssymb,
aps,
]{revtex4-1}

\usepackage{graphicx}
\usepackage{dcolumn}
\usepackage{bm}
\usepackage[dvipsnames]{xcolor}

\begin{document}                  

\title{Nanoscale heterogeneous dynamics probed by nanosecond x-ray speckle visibility spectroscopy}

\author{Yanwen Sun}
\affiliation{Linac Coherent Light Source, SLAC National Accelerator Laboratory, Menlo Park, California, 94025, USA}
\author{Gabriella Carini}
\thanks{Currently at Brookhaven National Laboratory, Upton, New York, 11973, USA}
\affiliation{Linac Coherent Light Source, SLAC National Accelerator Laboratory, Menlo Park, California, 94025, USA}
\author{Matthieu Chollet}
\affiliation{Linac Coherent Light Source, SLAC National Accelerator Laboratory, Menlo Park, California, 94025, USA}
\author{Franz-Josef Decker}
\affiliation{Linac Coherent Light Source, SLAC National Accelerator Laboratory, Menlo Park, California, 94025, USA}
\author{Mike Dunne}
\affiliation{Linac Coherent Light Source, SLAC National Accelerator Laboratory, Menlo Park, California, 94025, USA}
\author{Paul Fuoss}
\affiliation{Linac Coherent Light Source, SLAC National Accelerator Laboratory, Menlo Park, California, 94025, USA}
\author{Stephan O. Hruszkewycz}
\affiliation{Materials Science Division, Argonne National Laboratory, Lemont, 60439, U.S.A}
\author{Thomas J. Lane}
\affiliation{Linac Coherent Light Source, SLAC National Accelerator Laboratory, Menlo Park, California, 94025, USA}
\author{Kazutaka Nakahara}
\affiliation{Linac Coherent Light Source, SLAC National Accelerator Laboratory, Menlo Park, California, 94025, USA}
\author{Silke Nelson}
\affiliation{Linac Coherent Light Source, SLAC National Accelerator Laboratory, Menlo Park, California, 94025, USA}
\author{Aymeric Robert}
\affiliation{Linac Coherent Light Source, SLAC National Accelerator Laboratory, Menlo Park, California, 94025, USA}
\author{Takahiro Sato}
\affiliation{Linac Coherent Light Source, SLAC National Accelerator Laboratory, Menlo Park, California, 94025, USA}
\author{Sanghoon Song}
\affiliation{Linac Coherent Light Source, SLAC National Accelerator Laboratory, Menlo Park, California, 94025, USA}
\author{G. Brian Stephenson}
\affiliation{Materials Science Division, Argonne National Laboratory, Lemont, 60439, U.S.A}
\author{Mark Sutton}
\affiliation{Physics Department, McGill University, Montr\`eal, Quebec, Canada, H3A 2T8}
\affiliation{Linac Coherent Light Source, SLAC National Accelerator Laboratory, Menlo Park, California, 94025, USA}
\author{Tim B. Van Driel}
\affiliation{Linac Coherent Light Source, SLAC National Accelerator Laboratory, Menlo Park, California, 94025, USA}
\author{Clemens Weninger}
\thanks{Currently at MAX IV Laboratory, Lund, Sweden}
\affiliation{Linac Coherent Light Source, SLAC National Accelerator Laboratory, Menlo Park, California, 94025, USA}
\author{Diling Zhu}
\email{Send correspondence to: dlzhu@slac.stanford.edu}
\affiliation{Linac Coherent Light Source, SLAC National Accelerator Laboratory, Menlo Park, California, 94025, USA}
   
\date{\today}

\begin{abstract}
We report observations of nanosecond nanometer scale heterogeneous dynamics in a free flowing colloidal jet revealed by ultrafast x-ray speckle visibility spectroscopy. The nanosecond double-bunch mode of the Linac Coherent Light Source free electron laser enabled the production of pairs of femtosecond coherent hard x-ray pulses. By exploring the anisotropic summed speckle visibility which relates to the correlation functions, we are able to evaluate not only the average particle flow rate in a colloidal nanoparticle jet, but also the heterogeneous flow field within. The reported methodology presented here establishes the foundation for the study of nano- and atomic-scale heterogeneous fluctuations in complex matter using x-ray free electron laser sources.
\end{abstract}

\maketitle

Nanoscale fluctuations of matter are closely related to transport, polarization, and mechanical properties in a wide range of materials. Key examples currently under study include ferroelastic domains in relaxors~\cite{fu2009relaxor,krogstad2018relation}, martensitic transformations in shape memory alloys~\cite{sanborn2011direct}, and plastic deformation mechanisms in metallic glasses~\cite{berthier2011dynamic, evenson2015x, luo2017relaxation}.
Numerical simulations~\cite{berthier2011theoretical,li2016origin} predict that the dynamics in such systems is often spatially heterogeneous and temporally intermittent, e.g. through nanoscale `avalanches' of fast collective motion, rather than homogeneous diffusion of individual atoms.
Experimental studies using thermal, mechanical, or dielectric probes~\cite{wilde2002slow,wagner1999dielectric,qiao2013relaxation} can probe the temporal behavior but lacks nanoscale spatial resolution.
Other techniques like dynamic light scattering~\cite{ballesta2004temporal} are also limited to large length scales. Its analogue at x-ray wavelengths, x-ray photon correlation spectroscopy (XPCS), provides sensitivity on the nano and atomic length scale. However, XPCS studies have so far been limited to slow dynamics due to the relatively small scattering cross section from atomic order and the low coherent flux at current x-ray sources~\cite{shpyrko2014x}.

X-ray free electron lasers (FELs) provide a new playground for XPCS measurements with an unprecedented high coherent flux delivered within sub-100-femtosecond pulses~\cite{bostedt2016linac}. Each x-ray pulse captures a snapshot of the atomic arrangements of the system~\cite{stephenson2009x}. A pair of femtosecond pulses with a time separation in the femto- to nanosecond timescales enables the capability of resolving fluctuations at these much faster timescales~\cite{deckertwo,Osaka2016,roseker2009performance,lu2016design, zhu2017development,sun2019compact}, relevant to systems with avalanche behaviors. While such timescales are beyond the current time resolution of x-ray detectors, visibility spectroscopy that relies on analyzing the contrast change in the summed speckle has been proposed, providing information equivalent to intensity autocorrelation functions~\cite{shenoy2003lcls,gutt2009measuring}. Following the first observation of high contrast speckle from atomic-scale order at x-ray FELs~\cite{hruszkewycz2012high}, much progress has been made towards applying the two-pulse modes and visibility spectroscopy to investigate homogeneous dynamics in various material systems~\cite{roseker2018towards,seaberg2017nanosecond,esposito2020skyrmion,shinohara2020split}.
It has been proposed that heterogeneous dynamics can be characterized by measuring higher order spatio-temporal correlation functions~\cite{berthier2011dynamic,madsen2010beyond}.
In this paper, we show how heterogeneous dynamics can alternatively be analyzed through the dependence of the time correlation functions on the magnitude and direction of the scattering wavevector.
We explore nanosecond colloidal dynamics with a nanoscale non-uniformity imposed using a flowing liquid jet. The methodology of probing the anisotropy of the correlation functions demonstrated here provides a general pathway towards understanding heterogeneous dynamics and dynamical heterogeneity in disordered systems at previously inaccessible timescales. 

\begin{figure}
    \centering
    \includegraphics[width=1\linewidth]{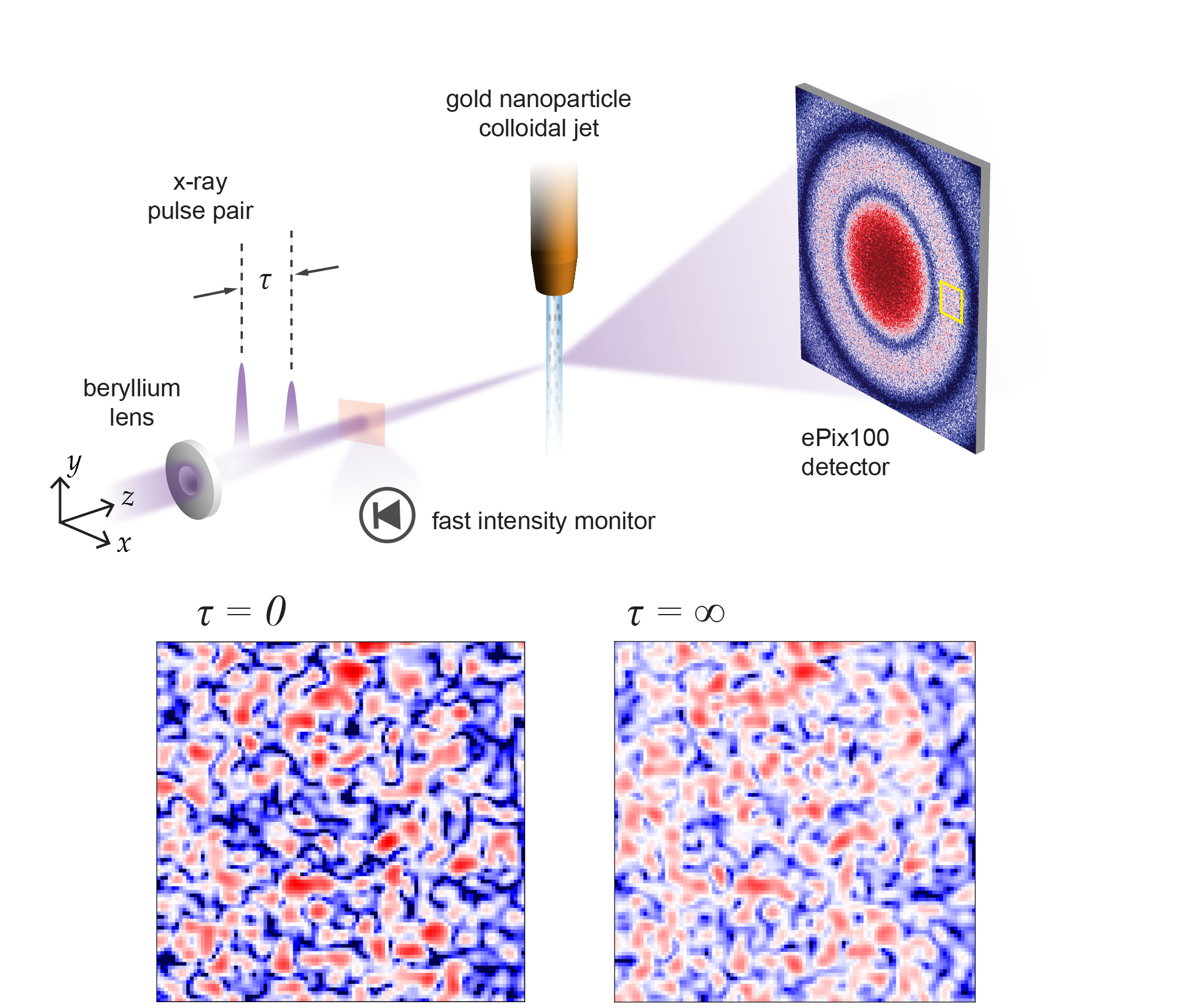}
    \caption{Schematic of the two-pulse XPCS experiments. Two x-ray pulses with a time separation of $\tau$ were generated using the nanosecond double-bunch mode and delivered to the sample. A high-speed intensity monitor upstream of the sample measured the relative intensities of the two pulses within each pulse pair. A 2D detector 8 meters downstream of the sample recorded the sum of the scattering from each pulse pair. Enlarged views are shown of the region of interest outlined by the yellow rectangle of the simulated speckle pattern sum for $\tau = 0$ and $\tau \rightarrow \infty$, illustrating the loss of contrast when $\tau$ exceeds the time scale of the dynamics.}
    \label{fig:schematics}
\end{figure}

The experiment was carried out at the x-ray correlation spectroscopy instrument at the Linac Coherent Light Source~\cite{alonso2015x} with the FEL operating in the so-called nanosecond double-bunch mode~\cite{deckertwo}. Pulse pairs separated by $\tau = 49$~ns were used.
The pulse pairs were were attenuated by a factor of 20 to avoid beam heating and monochromatized using a 4-bounce Si(111) monochromator at 8.2~keV, with an average total pulse energy of 0.03~$\mathrm{\mu J}$ measured at the sample plane.
Beryllium compound refractive lenses $\sim 400$~m downstream from the undulator focused the beam to $\sim 3~\mu m$ at the sample with a focal length of $\sim 3.3$~m. Slits at the lens limited the numerical aperture and provided a larger and more stable focal spot.
Figure~\ref{fig:schematics} shows the experimental schematic. The sample was a liquid water jet containing gold nanospheres of $R = 50$~nm radius (Nanopartz, 5~mg/ml, $\Phi$ = 0.026 vol$\%$, capped with carboxylic acids for stabilization).
The flow was adjustable using a Shimadzu liquid chromatography pump, and was delivered via a cylindrical glass capillary nozzle with $d_c = 100~\mu$m inner diameter.
Upon exiting the nozzle, the boundary condition change led to the shrinkage of the jet known as the \emph{vena-contracta} effect~\cite{haustein2017simple} and the diameter was measured to be $d = 92~\mu m$ at the x-ray interaction point $l_0 = 1.4$~mm below the nozzle using an optical microscope.
A transmissive high-speed intensity monitor upstream of the sample provided a measurement of the relative intensity of the pulses within each pulse pair~\cite{sun2018pulse}.
An ePix100 detector (pixel size, 50 $\mu m$, $704\times 768$ pixels) 8~m downstream of the sample measured the small angle scattering~\cite{carini2016epix100, sikorski2016application}, each exposure capturing the sum of x-ray scattering from a pulse pair at 120~Hz.

As proposed in Ref.~\cite{shenoy2003lcls,gutt2009measuring}, the speckle contrast (the normalized variance of the intensity distribution in the speckle pattern) of the sum was obtained. This is equivalent to the intensity correlation $g_2$ measured in sequential XPCS experiments~\cite{carnis2014demonstration,lehmkuhler2018dynamics}.
The scattering sum was recorded at flow rates between 1 to 12~mL/min, corresponding to an average speed $\bar{v} = 2.5 - 30.1$~m/s using the jet diameter of 92~$\mathrm{\mu m}$.
At 12~mL/min, the Reynolds number at the capillary exit was $\approx 2876$, at which the flow is generally considered in transition to the turbulent regime~\cite{avila2011onset}.
For flow rates from 1 to 8~mL/min, the flow was laminar. This agrees with observation of the jet changing from a clear gradually narrowing stream to a broadened hazy appearance at $\approx 14$~mL/min.

\begin{figure}[h!]
    \centering
    \includegraphics[width=0.8\linewidth]{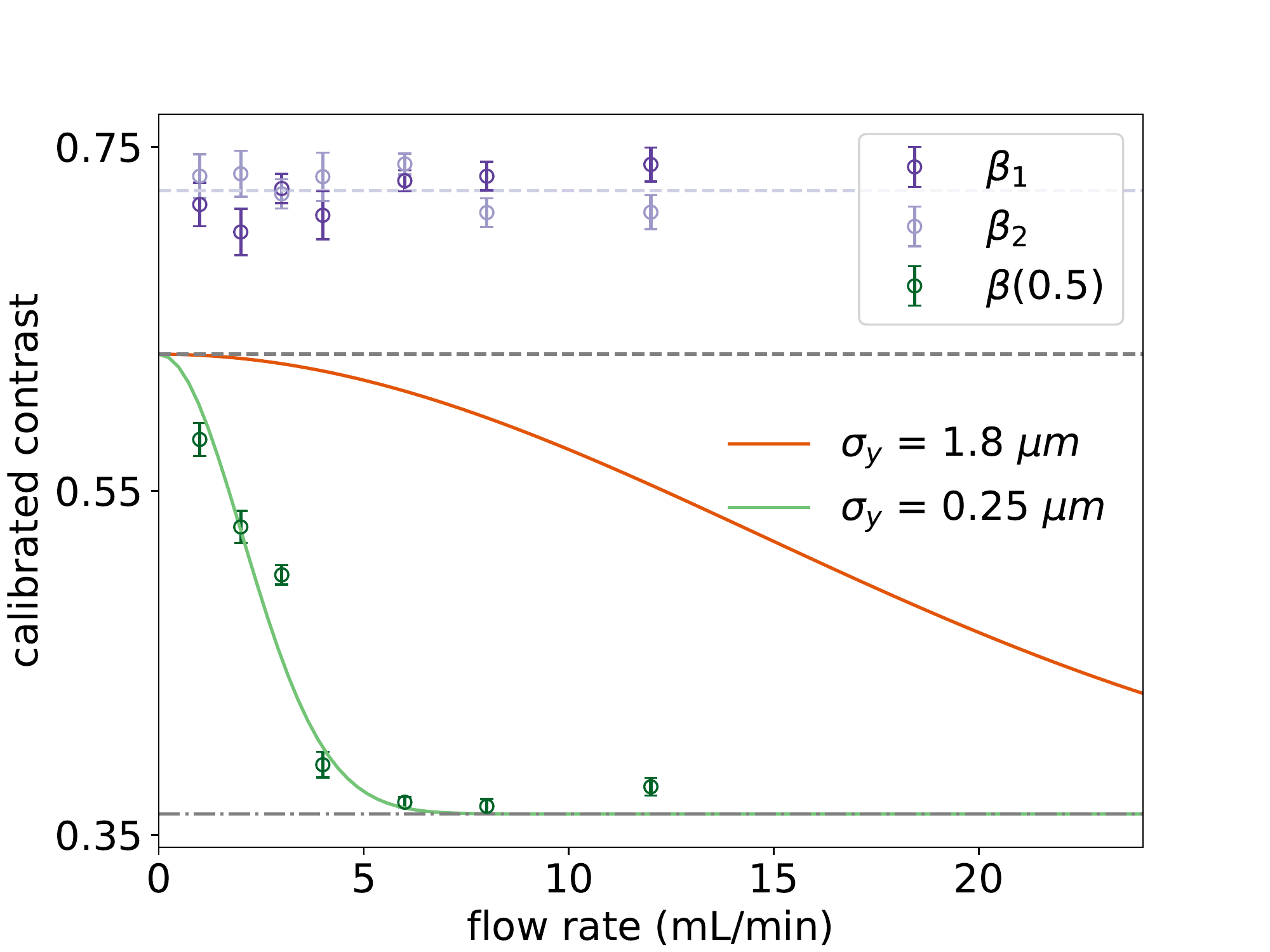}
    \caption{Calibrated contrasts for $r = 0, 0.5, 1$ as a function of flow rate. The curves show the calculated contrast decay assuming a uniform jet with different beam sizes. The two dashed gray lines indicate the range for $\beta(0.5)$. The high limit is smaller than $\beta_{1(2)}$due to the non-ideal spatial overlap between the two pulses.}
    \label{fig:wholeRing}
\end{figure}

Figure~\ref{fig:wholeRing} shows the measured speckle contrast (after calibration) as a function of the flow rate for an annular region of interest (ROI) of average radius $Q=0.0055$~\AA$^{-1}$ and width $\Delta Q = 0.0013$~\AA$^{-1}$ (see Fig.~4(a) in Supplemental Material).
Accurate speckle contrast evaluation requires several key calibration steps.
First, contrast reduction induced by sample dynamics must be separated from that due to x-ray source effects such as relative pulse pair intensity fluctuations and deviation from perfect spatial overlap.
The measured contrast $\beta$ is related to the intermediate scattering function $f(\bm{Q},\bm{v},\tau)$ (where $\bm{v}$ denotes the jet velocity profile) via 
\begin{equation}
  \beta = r^2\beta_1+(1-r)^2\beta_2+2r(1-r) \beta_1 \mu |f(\bm{Q},\bm{v},\tau)|^2.
  \label{eq:beta}
\end{equation}
Here $\beta_1$ and $\beta_2$ are the contrast values for each of the pulses in the pair.
The pulse intensity ratio $r \equiv i_1/(i_1+i_2)$ varies from pulse to pulse.
The parameter $\mu$ quantify the effective spatial overlap.
An estimate of $\mu\approx 0.74 \pm 0.02$ was obtained from analysis of scattering from a static reference sample (see Supplemental Material Section I).
The second calibration step addresses systematic statistical error of the contrast evaluation algorithms~\cite{sun2020accurate}, where biased output of the photon assignment algorithms that are detector-response dependent can be removed.
Moreover, mean intensity variation within the ROI can lead to an overestimation of the contrast.
Following the procedure in Supplemental Material Section II, we first grouped the scattering patterns based on $r$, extracted and corrected the contrasts for different $r$ values, and subsequently fit the corrected contrasts to Eq.~\ref{eq:beta} to extract $\beta_1$, $\beta_2$ and $|f|^2$. 
We confirmed that $\beta_1$ and $\beta_2$ (dark/light purple in Fig.~\ref{fig:wholeRing}) agreed with each other for all flow rates within the error bars, indicating stability of the setup. The contrasts for $r= 0.5$ were calculated and plotted as green circles in Fig.~\ref{fig:wholeRing}, where a rapid decrease was observed between 0 and 6~mL/min. 

\begin{figure}
    \centering
    \includegraphics[width=\linewidth]{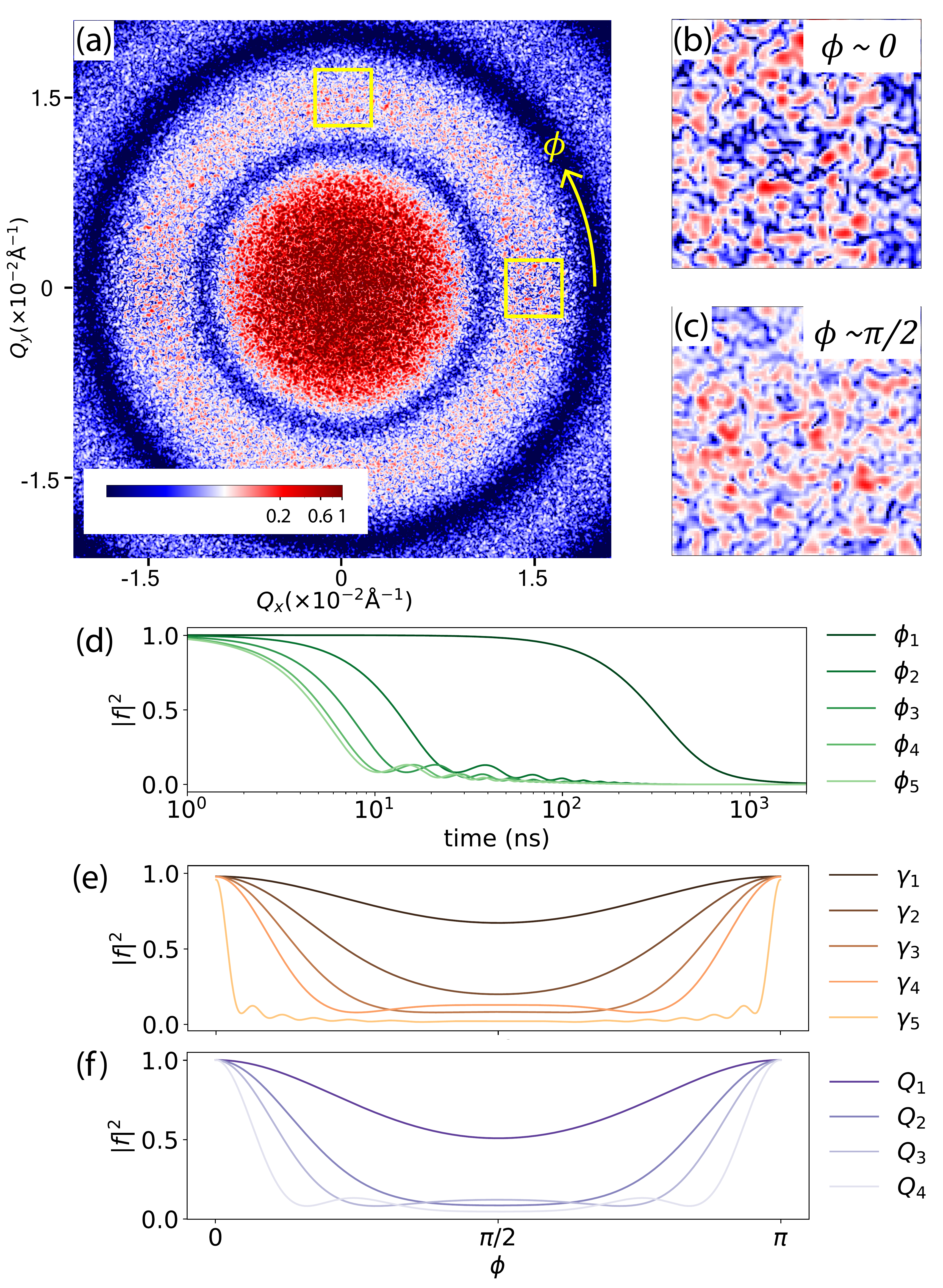}
    \caption{Simulation of the anisotropic jet dynamics. (a) Simulated speckle pattern sum from a parabolic-flow jet at an averaged speed of $\bar{v} = 0.5$~m/s. (b) and (c) are enlarged views of the two ROIs outlined in yellow centered at $0.015$~\AA$^{-1}$ for $\phi\sim 0$ and $\phi\sim \pi/2$. (d) For fixed $Q = 0.0055$~\AA$^{-1}$, $\gamma = 0$ (parabolic flow), and $\bar{v} = 5$~m/s, calculated $|f|^2$ as a function of time for $\phi_{1-5} =0, \pi/8, \pi/4, 3\pi/8$, and $\pi/2$. (e) For fixed $Q = 0.0055$~\AA$^{-1}$, $\bar{v} = 7.5$~m/s and $\tau = 49$~ns, calculated $|f|^2$ as a function of azimuthal angle $\phi$ for $\gamma_{1-5} = 0.95, 0.9, 0.85, 0.8$, and $0$. (f) For fixed $\bar{v} = 0.5$~m/s, $\gamma = 0$ (parabolic flow), and  $\tau = 49~$ns, calculated $|f|^2$ as a function of azimuthal angle $\phi$ for $Q_{1-4} = 0.0055, 0.011, 0.015$, and $0.024$~\AA$^{-1}$.}
    \label{fig:anisotropySimulation}
\end{figure}

The contrast decrease can be primarily attributed to the displacement of the gold nanospheres in flow. Free diffusion, has an estimated time scale of $\sim 70$~$\mu$s at this $Q$ range. Being much longer than the pulse separation, it can be ignored.
Therefore, the intermediate scattering function at the scattering vector $\bm{Q}$ is 
\begin{equation*}
\begin{split}
    f(\bm{Q},\bm{v}, \tau) & \approx \frac{1}{\langle N \rangle}\sum_{j=1}^{N(0)} \sum_{k=1}^{N(\tau)} \frac{\langle E_{j}E_{k} e^{i\bm{Q}\cdot[\bm{r}_k(\tau)-\bm{r}_j(0)]}\rangle}{\langle E_{j} E_{k}\rangle} \\
    & = \frac{1}{\langle N \rangle}\sum_{j=1}^{N(0)} \frac{\langle  E_{j}(\tau)E_{j}(0) e^{i\bm{Q}\cdot[\bm{r}_j(\tau)-\bm{r}_j(0)]}\rangle}{ \langle E_{j}^2\rangle},\\
\end{split}
\end{equation*}
approximating the particle displacement using the velocity field of the jet $\bm{r}_j(\tau)-\bm{r}_j(0) = \int_0^{\tau}\,\bm{v}dt$. Here $E_{j}(\tau)$ is the electric field amplitude on the $j$-th gold nanoparticle at time $\tau$. Since the large number of particles in the scattering volume ($\langle N \rangle \approx 3.2\times 10^2$) samples the electric field over a large number of pulse pairs, the summation can be approximated as the integration over the illumination volume
\begin{equation*}
    f(\bm{Q},\bm{v}, \tau)= \frac{ \int_{V_0} E(\bm{r}',\tau)E(\bm{r}, 0)\exp(i\bm{Q}\cdot \bm{v} \tau) d\bm{r}}{\int_{V_0}[E(\bm{r},0)]^2d\bm{r}}.
\end{equation*}
Here $\bm{r}$ is the location in the sample at time $0$. At time $\tau$, the new location is at $\bm{r'} = \bm{r}+\int_0^\tau  \bm{v} dt \approx \bm{r} + \bm{v}\tau$. $E(\bm{r},\tau)$ is then the electric field amplitude at time $\tau$ for the part of the sample at location $\bm{r}$ at time 0.

We first consider a uniform speed distribution, i.e., all gold particles have the same speed ($\bm{v} = -\bar{v}\hat{\bm{y}}$).
For a Gaussian beam spatial profile 
$$
E(\bm{r},\tau) \approx E_0 \exp[-x^2/(2\sigma_x^2)] \exp[-(y-\bar{v}\tau)^2 /(2\sigma_y^2)],
$$
Here we neglect the z-dependence of the electric field as the sample thickness is much smaller than the Rayleigh length of the x-ray beam. See Supplemental Material Section III for more details.
The intermediate scattering function reduces to
\begin{equation}
        |f(\bm{Q},\bar{v}, \tau)| = \exp[-\frac{(\bar{v}\tau)^2}{4\sigma_y^2}].
\label{eq:intermediateScatteringuniform}
\end{equation}
In this case the result is independent of $\bm{Q}$, since there is no spatial structure to the dynamics.
A best fit (green line in Fig.~\ref{fig:wholeRing}) yields an x-ray beam size estimation of $\sigma_y \sim 0.25~\mu$m.
This significantly deviates from $\sigma_y \sim 1.8~\mu$m determined from the speckle size from a static reference scattering sample.
Clearly the uniform-flow model is an oversimplification of the particle dynamics within the jet at $\tau = 49$~ns. A non-uniform-flow model is required to more accurately describe our observations~\cite{lhermitte2017velocity}. 

We now show that the fast decay originated from the circular average over the anisotropic behavior of the decay. It is well known that when viscous liquid enters a pipe, given sufficient distance, a parabolic pipe flow profile forms.
Considering the flow rate used in our experiment of 1-12~mL/min, and the capillary length of $\sim$ 20 mm, this was the case. 
As the fluids exit the capillary, the boundary condition imposed by the wall of capillary is lifted.
The radial speed difference will gradually become smaller as the center slows down and the outer part of the jet speeds up.
We model the flow profile as a linear combination of the uniform and parabolic components, and use $\gamma$ ($0 \le \gamma \le 1$) to indicate the fraction of the uniform flow component, such that the speed field can be written as
\begin{equation}
v(x,z) = \gamma\bar{v} + 2(1-\gamma)\bar{v} [1-\frac{4(x^2+z^2)}{d^2}].
\end{equation}
The speed at the center of the jet is $(2-\gamma) \bar{v}$ and gradually decreases towards the boundary to $\gamma \bar{v}$. 
Considering a Gaussian beam and since $\sigma_x/d \ll 1$, we neglect all the $O((\sigma_x/d)^2)$ terms. The intermediate scattering function takes the form
\begin{equation}
\begin{split}
  |f(\bm{Q},v, \tau)|  &\approx  \frac{1}{2}|\int_{-1}^{1} \exp[i\frac{\tau}{\tau_2}2(1-\gamma)\zeta^2]\times\\ &\exp\{-\frac{\tau^2}{\tau_1^2}[2(1-\gamma)(1-\zeta^2)+\gamma]^2\} ~d\zeta|,
  \label{eq:phiDependence}
  \end{split}
\end{equation}
which introduces an additional timescale $\tau_2 = [Q\bar{v}\sin\phi]^{-1}$, with a dependence on $\phi$ defined as the angle between $\bm{Q}$ in plane and $\hat{\bm{x}}$.
The negative exponential factor dependent on $\tau_1 = 2\sigma_y/\bar{v}$ characterizes the decrease of the intermediate scattering function due to the averaged displacement of the jet.
The phase factor dependent on $\tau_2$ modulates this amplitude. It reflects the effect of the velocity gradient within the jet.
At $\phi = 0$, the phase factor is unity. 
At $\phi = \pi/2$, the modulation term is significant as $\tau/\tau_2 $ varies from 6.7 to 81 for flow rates 1-12 mL/min.
To better illustrate the $\phi$ dependence, we use simulated speckle patterns from particles in a parabolic flow profile (see Appendix IV), which are displayed in Fig.~\ref{fig:anisotropySimulation}(a-c). They show the sum of scattering from two instantaneous particle positions separated in time.
One can see a clear speckle visibility difference at different angular positions along a ring of constant $|\bm{Q}|$.
Similar observations of the anisotropic correlation functions were reported in the XPCS measurement of stress/strain relaxation in amorphous materials like colloidal glasses and polymers~\cite{dallari2020microscopic,sutton2020high}. Figure~\ref{fig:anisotropySimulation}(f) shows $|f|^2$ as a function of $\phi$, revealing a strong dependence in the vicinity of $\phi\sim 0$ or $\pi$. 
This dependence is most prominent at small $\gamma$ values. 
The timescales at $\phi_1 = 0$ and $\phi_5 = \pi/2$ differ by up to 2 order of magnitude  as shown in Fig.~\ref{fig:anisotropySimulation}(d).
Moreover, as plotted in Fig.~\ref{fig:anisotropySimulation}(e), one can see that $|f(\phi)|^2$ is sensitive to $\gamma$, e.g., a 5\% decrease from $\gamma_1 = 0.95$ to $\gamma_2 = 0.9$ leads to $|f|^2$ decreasing from 0.67 to 0.20 at $\phi = \pi/2$. 

\begin{figure}
    \centering
    \includegraphics[width=\linewidth]{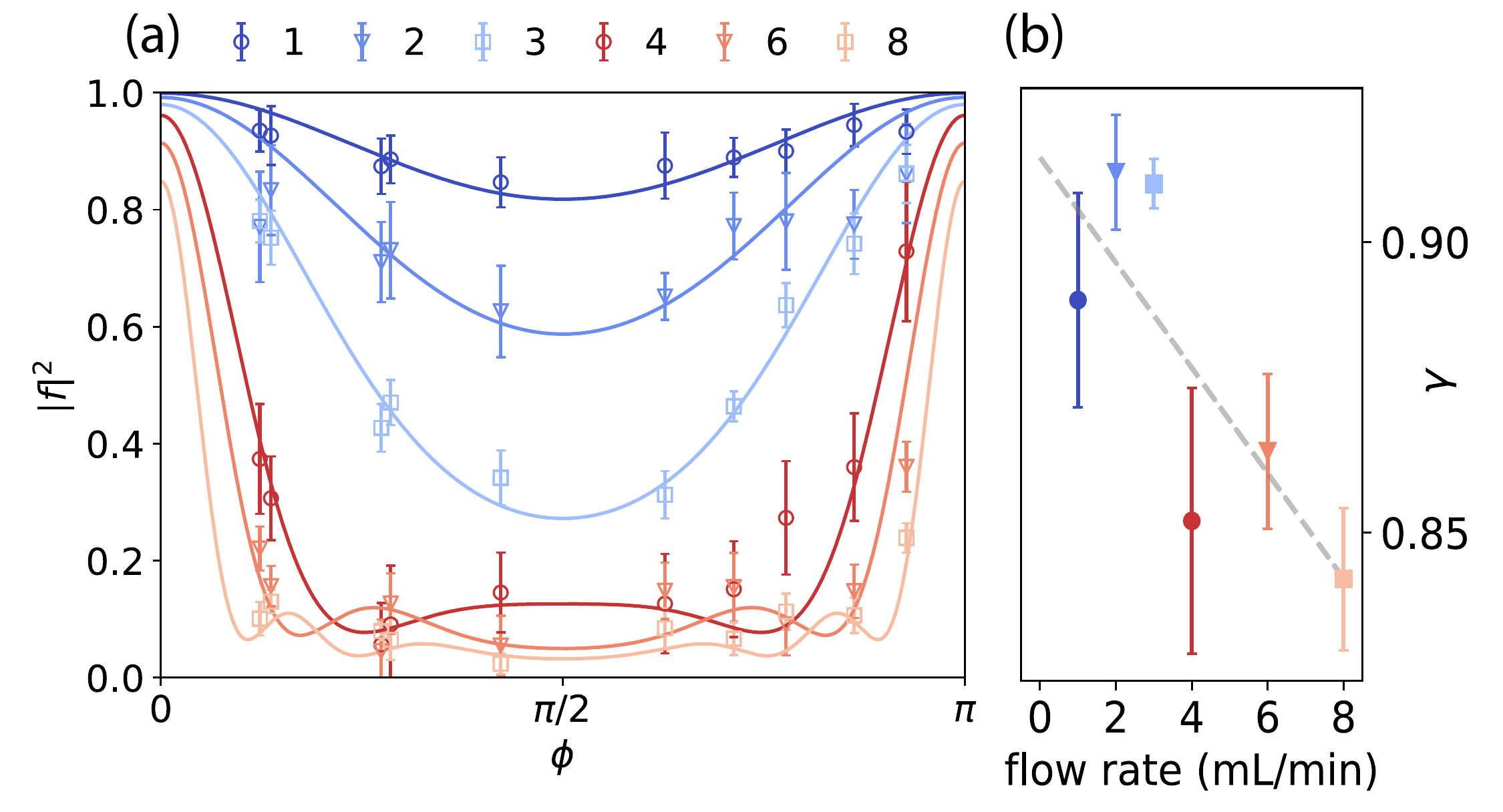}
    \caption{ (a) Dependence of $|f|^2$ on azimuthal angle $\phi$ for six flow rates from 1 to 8~mL/min, revealing anisotropic dynamics. The curves are fits giving the $\gamma$ values displayed in (b). The dashed line is a guide to the eye.}
    \label{fig:anisotropyExp}
\end{figure}

To analyze the $\phi$ dependence in the data, the annular ROI is divided into 10 sectors, each covering $\delta \phi \approx 0.15 \pi$. The $\phi$ value of each sector is defined using its centroid, reduced to the range of $[0, \pi)$ assuming the equivalency between the scattering in $\phi$ and $\phi+\pi$.
The intensity of the intermediate scattering function is plotted for these $\phi$ regions in the laminar flow cases with flow rates from 1 to 8~mL/min in Fig.~\ref{fig:anisotropyExp}(a). For each flow rate, we numerically calculated the optimal $\gamma$ values by least squares, which is displayed in Fig.~\ref{fig:anisotropyExp}(b). 
One can see a $\sim 5\%$ reduction in the $\gamma$ values as the flow rate increases.
This qualitatively matches the notion of the effective jet flow length $l_e = l_0/(d_c \mathrm{Re})$~\cite{haustein2017simple}, which quantifies the evolution of the velocity profile. Smaller $l_e$ indicates an earlier stage in the profile transition. 
With a fixed beam sample interaction location, the travel length $l_0$ in free-flight is fixed, whereas it decreases as the flow rates increases giving a less uniform flow and a smaller $\gamma$.

Our result presents the first observation of dynamic visibility anisotropy at an x-ray FEL. We have shown that velocity profiles can be evaluated in detail, which has many applications~\cite{fuller1980measurement,narayanan1997measurement,lhermitte2017velocity}.
From the pulse pairs in which one of them dominate in total intensity, the high contrast indicates that a single pulse `freezes' the motion. 
Thus by using pulse pairs with shorter time separations including the nanosecond double-bunch mode or the split-delay systems~\cite{deckertwo,Osaka2016,roseker2009performance,sun2019compact}, one can explore the turbulent regimes.
One surprising yet general implication of our result relates to the highly anticipated experiments aiming at the study of atomic-scale dynamics of supercooled liquids with the upcoming high repetition rate x-ray FELs.
In order to refresh the sample for each probe pulse pair, the samples need to be delivered at high speed, either via droplets or jets.
One must make sure the $\tau_2$ equivalent of the jet delivery mechanism does not introduce dynamics on the same time scale as the intrinsic dynamics of the sample.
Take a laminar water jet for example, with an average speed of 20-50 m/s, $\tau_2$ will be in the ps time scale, e.g. at the structure factor maximum near 2~$\mathrm{\AA}^{-1}$.
The internal collective flow within high speed micro droplets could be in the ps time scale as well. In such cases, $\phi$ dependent contrast analysis will be mandatory in order to quantitatively isolate the relevant dynamics information.

On the other hand, our observation also demonstrates the radial modulation of speckle contrast as a probe sensitive to the velocity gradient and thus the size of the nanoscale dynamic regions in disordered systems.
By going to higher $Q$, the experimental observation and measurement protocol of the anisotropic nanoscale dynamics extends naturally to the atomic scale.
This technique can thus be exploited to provide a detailed view of the heterogeneous nature of disordered systems beyond simplistic time scale analysis and towards the evaluation of higher order correlation functions.

\begin{acknowledgements}
We thank Diego H. Villeneuve for helpful discussions. This work is supported by the U.S. Department of Energy, Office of Science, Office of Basic Energy Sciences under Contract No. DE-AC02-76SF00515. S.O.H. and G.B.S. supported by DOE Office of Science, Basic Energy Sciences, Division of Materials Science and Engineering.
\end{acknowledgements}

\bibliography{reference}
\end{document}



\title{Supplemental material for ``Nanoscale heterogeneous dynamics probed by nanosecond x-ray speckle visibility spectroscopy"}

\author{Yanwen Sun}
\affiliation{Linac Coherent Light Source, SLAC National Accelerator Laboratory, Menlo Park, California, 94025, USA}
\author{Gabriella Carini}
\thanks{Currently at Brookhaven National Laboratory, Upton, New York, 11973, USA}
\affiliation{Linac Coherent Light Source, SLAC National Accelerator Laboratory, Menlo Park, California, 94025, USA}
\author{Matthieu Chollet}
\affiliation{Linac Coherent Light Source, SLAC National Accelerator Laboratory, Menlo Park, California, 94025, USA}
\author{Franz-Josef Decker}
\affiliation{Linac Coherent Light Source, SLAC National Accelerator Laboratory, Menlo Park, California, 94025, USA}
\author{Mike Dunne}
\affiliation{Linac Coherent Light Source, SLAC National Accelerator Laboratory, Menlo Park, California, 94025, USA}
\author{Paul Fuoss}
\affiliation{Linac Coherent Light Source, SLAC National Accelerator Laboratory, Menlo Park, California, 94025, USA}
\author{Stephan O. Hruszkewycz}
\affiliation{Materials Science Division, Argonne National Laboratory, Lemont, 60439, U.S.A}
\author{Thomas J. Lane}
\affiliation{Linac Coherent Light Source, SLAC National Accelerator Laboratory, Menlo Park, California, 94025, USA}
\author{Kazutaka Nakahara}
\affiliation{Linac Coherent Light Source, SLAC National Accelerator Laboratory, Menlo Park, California, 94025, USA}
\author{Silke Nelson}
\affiliation{Linac Coherent Light Source, SLAC National Accelerator Laboratory, Menlo Park, California, 94025, USA}
\author{Aymeric Robert}
\affiliation{Linac Coherent Light Source, SLAC National Accelerator Laboratory, Menlo Park, California, 94025, USA}
\author{Takahiro Sato}
\affiliation{Linac Coherent Light Source, SLAC National Accelerator Laboratory, Menlo Park, California, 94025, USA}
\author{Sanghoon Song}
\affiliation{Linac Coherent Light Source, SLAC National Accelerator Laboratory, Menlo Park, California, 94025, USA}
\author{G. Brian Stephenson}
\affiliation{Materials Science Division, Argonne National Laboratory, Lemont, 60439, U.S.A}
\author{Mark Sutton}
\affiliation{Physics Department, McGill University, Montr\`eal, Quebec, Canada, H3A 2T8}
\affiliation{Linac Coherent Light Source, SLAC National Accelerator Laboratory, Menlo Park, California, 94025, USA}
\author{Tim B. Van Driel}
\affiliation{Linac Coherent Light Source, SLAC National Accelerator Laboratory, Menlo Park, California, 94025, USA}
\author{Clemens Weninger}
\thanks{Currently at MAX IV Laboratory, Lund, Sweden}
\affiliation{Linac Coherent Light Source, SLAC National Accelerator Laboratory, Menlo Park, California, 94025, USA}
\author{Diling Zhu}
\email{Send correspondence to: dlzhu@slac.stanford.edu}
\affiliation{Linac Coherent Light Source, SLAC National Accelerator Laboratory, Menlo Park, California, 94025, USA}
   
\date{\today}

\maketitle

\section{I. Characterization of the two-pulse properties}
\label{appendix:characterization}

 \begin{figure}
    \centering
    \includegraphics[width=0.5\linewidth]{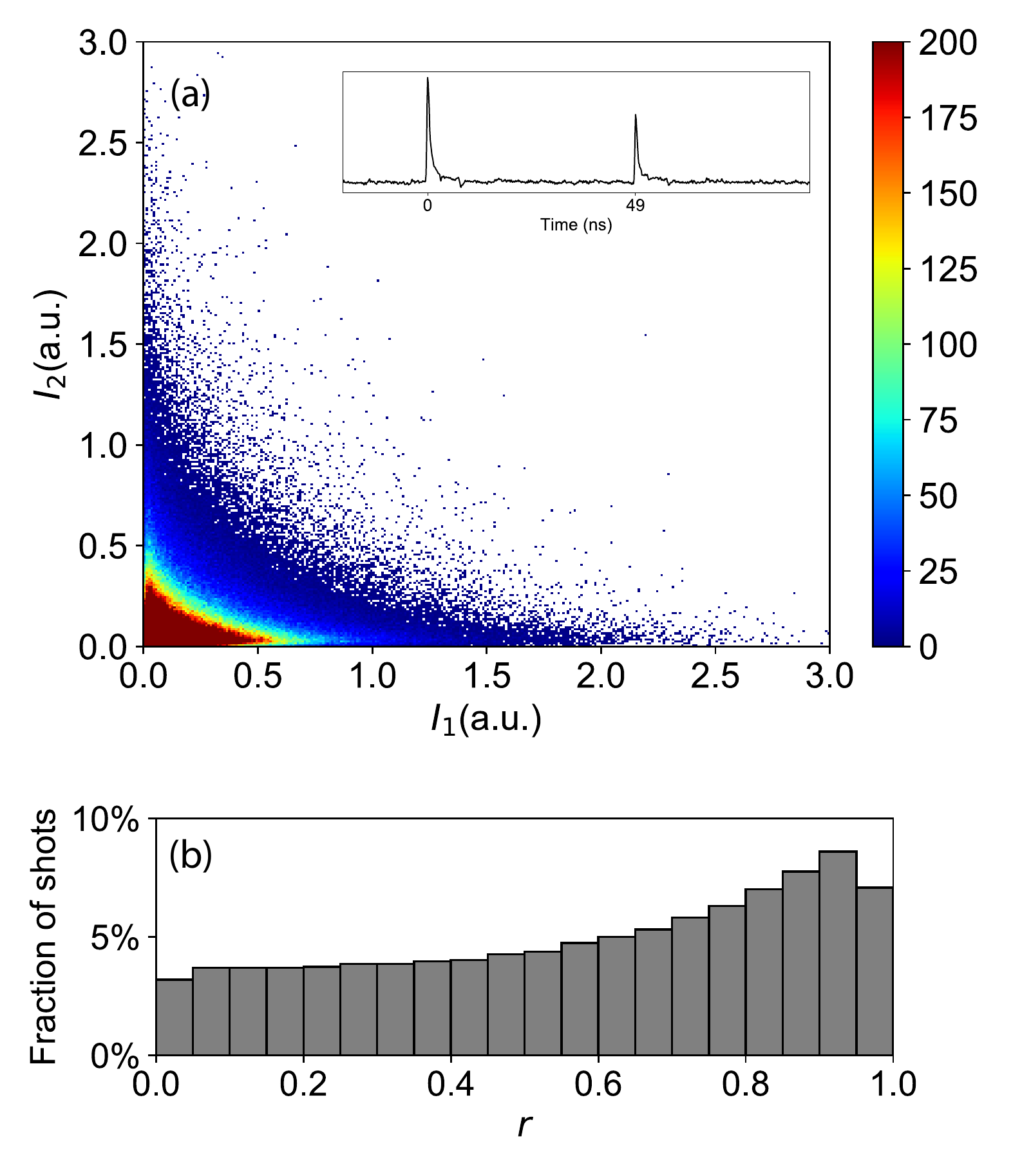}
    \caption{(Supplemental) (a) 2D cumulative histogram of the first pulse ($i_1$) and second pulse ($i_2$) intensity in a series of half a million shots measured using the transmissive intensity monitor. Inset: raw trace of a shot containing two pulses. (b) Histogram of the first pulse intensity fraction using the same data as in (a). }
    \label{fig:intensityDiagnostic}
\end{figure}

 \begin{figure*}
    \centering
    \includegraphics[width=1\textwidth]{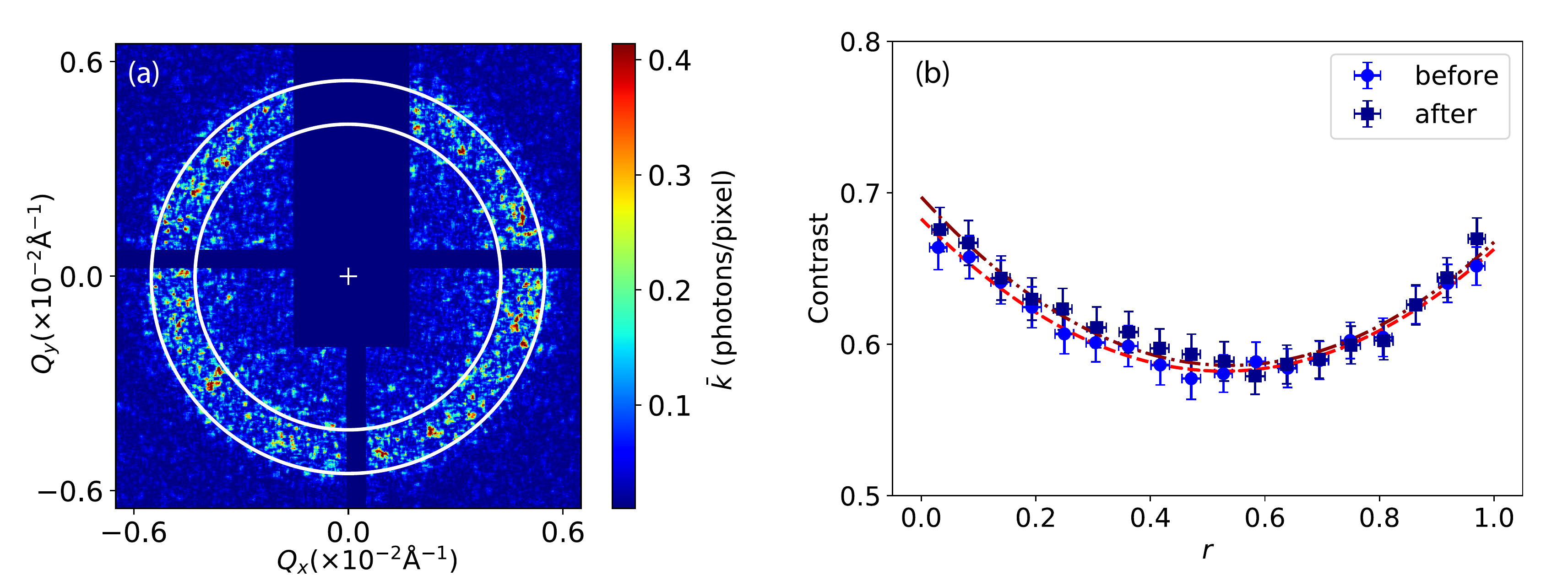}
    \caption{(Supplemental) (a) Small angle scattering from static silica powder. Displayed here is a zoomed in view of the averaged speckle pattern measured by the ePix100 detector. The cross indicates $Q=0$ and the beam center, shadowed in the beamstop. The dark blue areas indicate masked pixels that are not used in the data analysis. The annulus area outlined in white indicates an iso-$Q$ region that is subsequently used for analysis. (b) Extracted contrasts for different first pulse fractions.}
    \label{fig:overlap}
\end{figure*}

\begin{figure}[h!]
    \centering
    \includegraphics[width=0.5\linewidth]{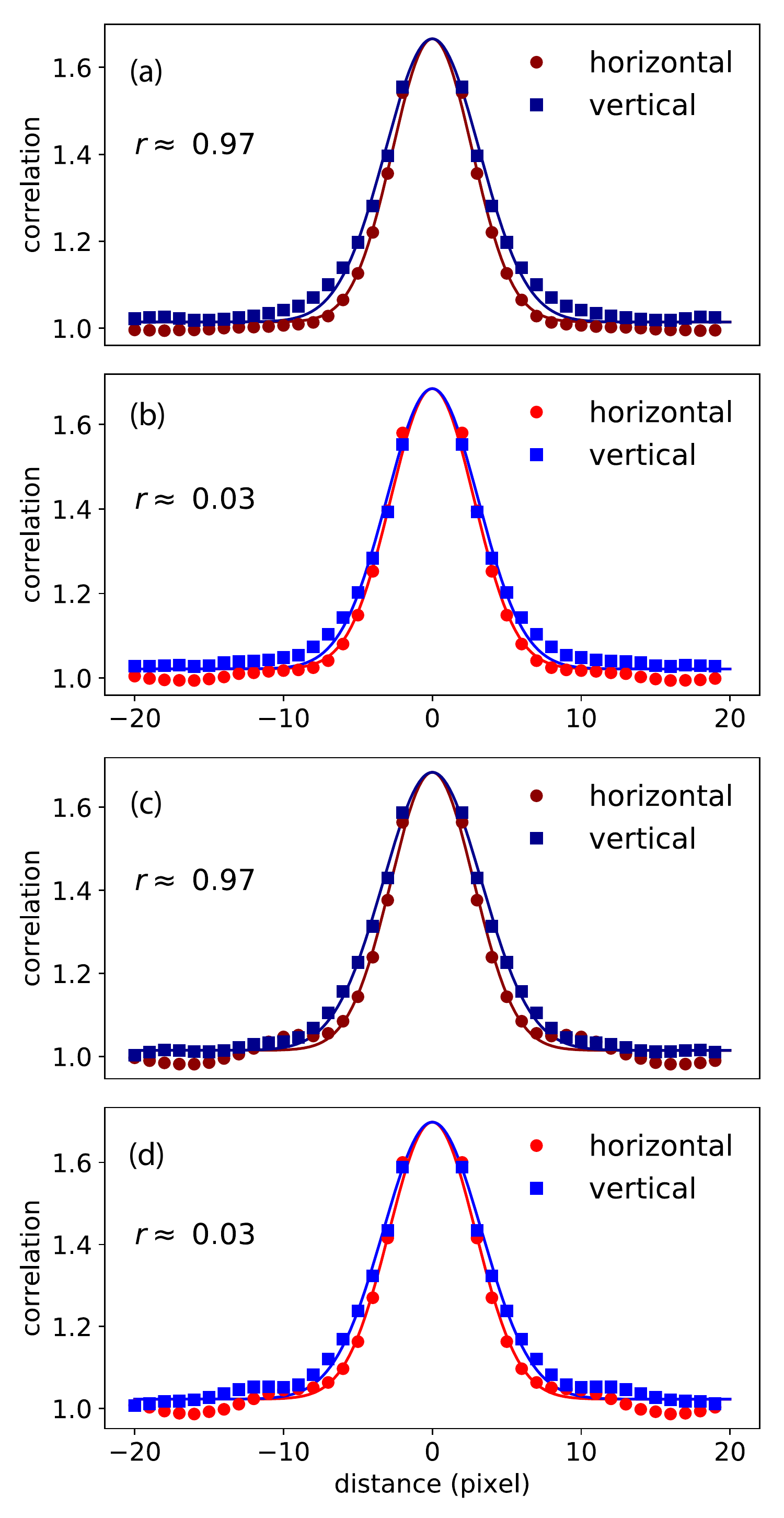}
    \caption{(Supplemental) Contrast extraction and speckle size analysis using the spatial auto-correlation of the scattering in the $Q$ region outlined in white ($0.0045-0.006$~\AA$^{-1}$) in Fig.~\ref{fig:overlap}(a) (Suppl.) before (a,b) and after (c,d) the jet flow measurement. The horizontal and vertical line-cuts at pixel-pair distances up to 20 pixels are plotted using red circles and blue squares separately. The curves are fitting results using a 2D Gaussian model with the fitted beam size (FWHM) along horizontal and vertical displayed in Table~I (Suppl.).}
    \label{fig:overlap_details}
\end{figure}
The nanosecond two-bunch mode is generated using two independent injector laser pulses. They are both synchronized to the Radio Frequency (RF) of the accelerator and strike the photo-cathode with a time separation precisely a multiple of the accelerating field (RF bucket). For LCLS this corresponds to a separation time by increments of 0.35~ns. The two electron bunches are then accelerated to their nominal energy and introduced in the undulator to emit x-rays independently via the self-amplified spontaneous emission (SASE) process. For the paper, we used the two-pulse time separation $\tau$ = 49~ns (140 RF buckets). The intensity diagnostic upstream the sample is able to temporally resolve the two pulses within a shot (see Fig.~\ref{fig:intensityDiagnostic}(a) (Suppl.) inset) and thus monitors the relative intensity of the first ($i_1$) and second ($i_2$) pulse on the sample shot-to-shot. The rather broad distribution shown in Fig.~\ref{fig:intensityDiagnostic}(a) (Suppl.) can be accounted for by the intrinsic fluctuations of the SASE process. We define $r$, which is the fraction of the first pulse intensity within a pulse pair, to label each frame recorded by the detector:
$$r = \frac{i_1}{i_1+i_2}, 0 \le r \le 1,$$ 
and plotted in Fig.~\ref{fig:intensityDiagnostic}(b) (Suppl.) is its histogram using the same data as in Fig.~\ref{fig:intensityDiagnostic}(a) (Suppl.).

\begin{table*}
\label{table:overlap}
\begin{tabular}{|c|c|c|c|c|c|c|c|c|}
    \hline
    time & $\beta_1$ & $\beta_2$ &  $\mu$ & $w_{x,1}~(\mu m)$ & $w_{y,1}~(\mu m)$ & $w_{x,2}~(\mu m)$ &  $w_{y,2}~(\mu m)$ \\ 
      \hline
  before  & $0.662\pm0.003$ & $0.682\pm0.003$ &$0.74\pm0.01$ & $3.42 \pm 0.07$ &$2.90\pm 0.06$ & $3.28 \pm 0.08$ & $2.99\pm0.07$\\
    \hline
  after & $0.667\pm0.005$ & $0.697\pm0.005$&$0.74\pm0.02$ & $3.29 \pm 0.07$ &$2.77\pm 0.06$ & $3.17 \pm 0.08$ & $2.76\pm0.06$ \\
  \hline
\end{tabular}
\caption{(Supplemental) Beam transverse coherence, spatial profile and overlap evaluation.}
\end{table*}

Besides intensity, the SASE process also introduces fluctuations in other pulse properties, e.g., coherence, trajectory, etc, leading to the dissimilarities between the two pulses delivered in each shot. And $\mu$ is introduced to quantify the contrast degradation due to these effects. For small angle scattering, $\mu$ represents the effective spatial overlap between the two pulses. For perfect spatial overlap, we have $\mu = 1$ as well as $\beta_1 = \beta_2$. $f(\bm{Q},\bm{v},\tau)$ is the intermediates scattering function denoting the contrast degradation resulting from the flow of the jet, which is a function of the momentum transfer $\bm{Q}$, the jet velocity distribution $\bm{v}$ and two-pulse time separation $\tau$. Using static sample, i.e., sample with correlation time much longer than the time separation $\tau$, the modulus of the intermediate scattering function $\approx 1$, and thus we can examine the contrast degradation due to the beam transverse coherence, the imperfect spatial overlap as well as finite detector sampling by analyzing the pulse pair contrast values as a function of $r$. As shown in Fig.~\ref{fig:intensityDiagnostic}(a) (Suppl.), the rather broad distribution of pulse-pair intensity also favors such investigation. 

A static silica nanoparticle sample (nominal radius: 50~nm, Kisker Biotech) was utilized for spatial overlap examination both before and after our jet flow measurement 12 hours apart. Shown in Fig.~\ref{fig:overlap}(a) (Suppl.) is its averaged scattering pattern over 1000 consecutive frames, where we can see visible speckles decorating the scattering due to purposely oversampling. The annulus outlined in white is the ROI selected for contrast analysis, where the averaged count rate is approximately 0.1 photons/pixel. Based on the values of $r$, the speckle patterns are grouped into different categories, from which we obtain the contrast values separately. Under this scenario, the contrast values can be extracted from the weighted average of the auto-correlations of the single-shot speckle patterns (weighted over the squared scattering intensity) grouped by $r$. Moreover, the sizes of the first and second pulses at the sample plane can also be extracted from the widths of the spatial auto-correlations in the case of $r\approx  1$ and $r\approx 0$. The line-cuts of the auto-correlations along the horizontal and vertical directions are plotted in Fig.~\ref{fig:overlap_details} (Suppl.) for the two cases both before (c, d) and after (e, f) the jet flow measurement 12 hours apart. The center 3 by 3 pixels in the auto-correlations are discarded due to Poisson noise and charge sharing between pixels. Thanks to the sufficient oversampling, we can still use 2D Gaussian fits to extract the contrast values (correlation values at zero distance) and the rms speckle size $S_{x(y)}$ (width of the gaussian distribution) measured by the detector. Moreover, considering finite detector sampling, given a pixel size $p = 50~\mu m$, the corrected rms speckle size is 
$$S'_{x(y)} = \sqrt{S_{x(y)}^2-(\frac{p}{\epsilon})^2}$$ with $\epsilon = \sqrt{6}$~\cite{sandy1999design}. As the speckles in the reciprocal space are the Fourier transform of the diffraction volume of the sample, in the case where the FEL beam is close to fully tranversely coherent, the speckle size $S'_{x(y)}$ and the illumination size on sample $w_{x(y)}$ follows~\cite{sikorski2015focus}
$$
w_{x(y)} = \frac{2\sqrt{2\log_e2}}{2\pi} \frac{\lambda L}{S'_{x(y)}}.
$$
Here $w_{x(y)}$ is the FWHM of the focal spot size and is related to the rms size of the electric field in the manuscript with $\sigma_{x(y)} = w_{x(y)}/(2\sqrt{\log_e2}) $ and $L = 8$~m is the sample-to-detector distance. 
We can calculate the beam size on the sample for the two pulses independently from the two extreme cases with $r \approx 0.97$ and $r\approx 0.03$. Moreover, as displayed in Fig.~\ref{fig:overlap}(b) (Suppl.), the contrasts obtained following this procedure for each $r$ group are plotted, with the curves representing the best fits to Eq.~1 of the main paper. The fitted values $\beta_1$, $\beta_2$ and $\mu$, together with the pulse-pair foci sizes are displayed in Table~I for the two 
measurement time points. The values are all the same within the error bars, indicating good overall stability of the pulse-pair properties during the course of the measurement.

\section{II. Contrast correction}
\label{appendix: correction}
\subsection{Non-uniform intensity in the ROI}
\begin{figure*}
    \centering
    \includegraphics[width=0.9\linewidth]{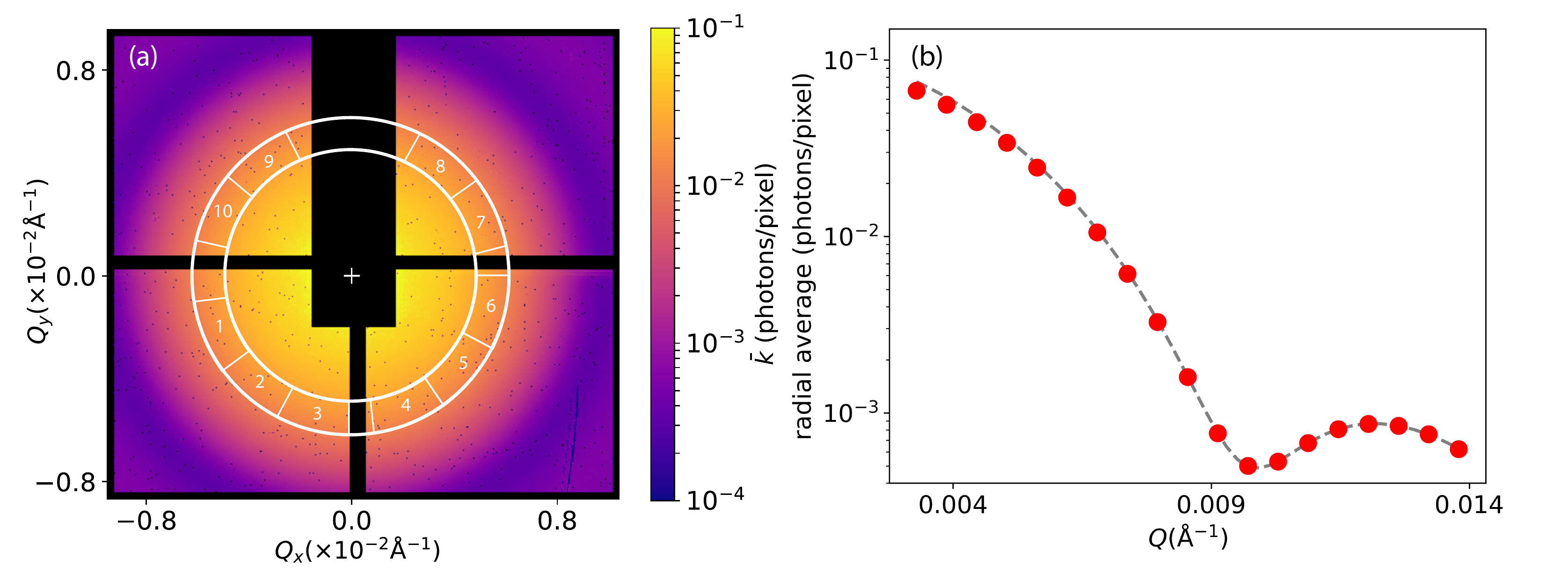}
    \caption{(Supplemental) (a) An averaged photon map from 1.3 million images measured by the ePix100 detector. The pixels masked in black showed abnormal behaviors and were not used in the data analysis. The annulus outlined by white with averaged $Q\approx 0.0055$~\AA$^{-1}$ was chosen for the analysis. The 10 sectors marked as 1 to 10 are the 10 regions for $\phi$ dependent analysis. (b) Radial average of the averaged scattering in (a). The dashed line is a fit to the sphere form factor assuming a Gaussian polydispersity, giving an average radius $\bar{R} = 47.5 \pm 0.5$~nm and rms polydispersity $\sigma_R \approx 9\% \pm 1\%$.}
    \label{fig:ave_scattering}
\end{figure*}

\begin{figure*}
    \centering
    \includegraphics[width=0.5\linewidth]{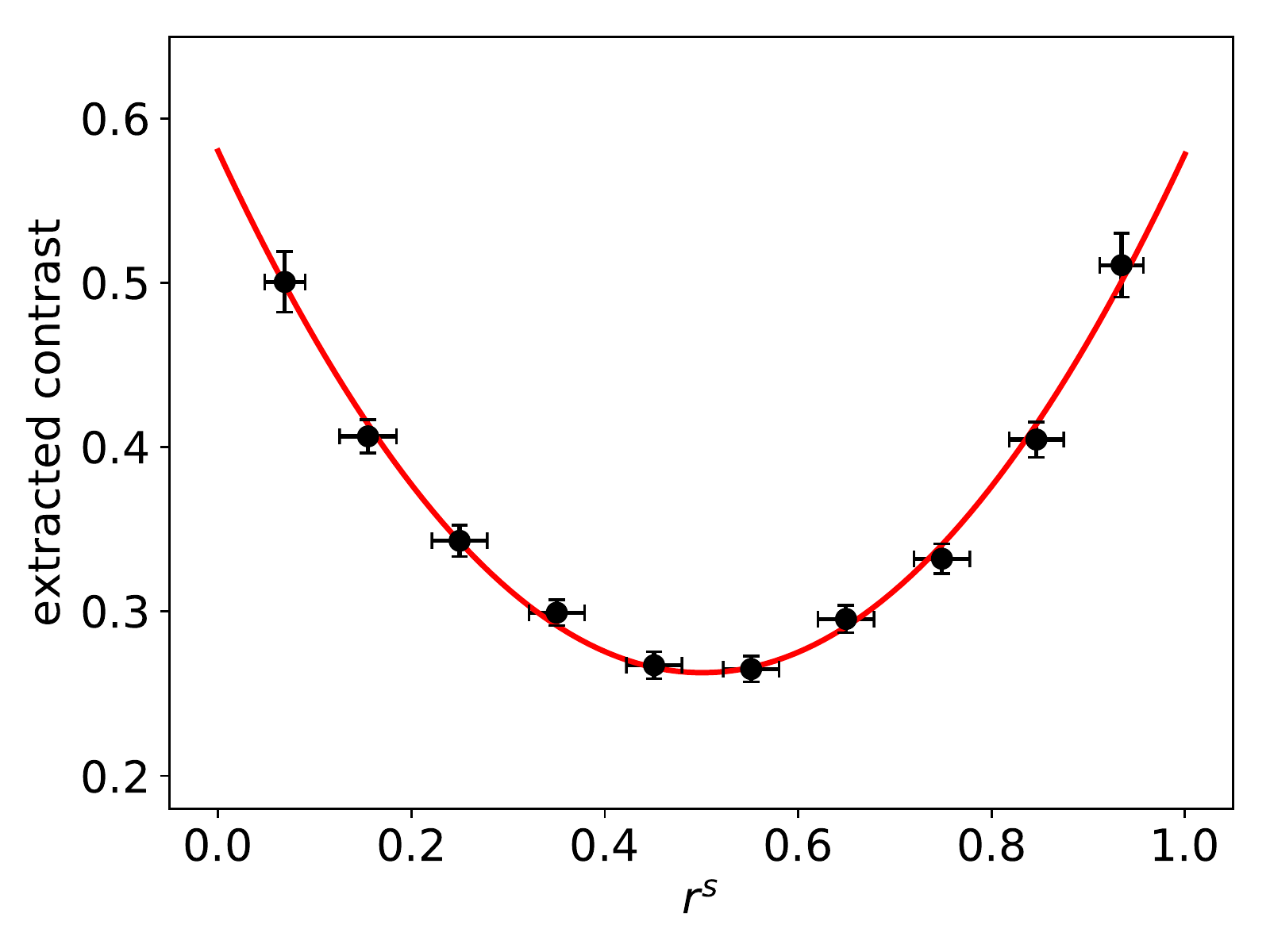}
    \caption{(Supplemental) Extracted contrasts of the artificial summed-scatterings at different first pulse-pair intensity fractions in the ring ROI. Red curve is a fit to the extracted contrasts using Eq.~1 of the main paper, and it yields a $\beta_e^s(r^s= 0)=  0.58 \pm 0.01$, $\beta_e^s(r^s = 1) =  0.58 \pm 0.01$ and $\mu_e = -0.10 \pm 0.02$.}
    \label{fig:contrast_correction}
\end{figure*}

Shown in Fig.~\ref{fig:ave_scattering}(a) (Suppl.) is the averaged scattering from the colloidal jet from 1.3 millions frames recorded by the ePix100 detector. The black pixels are discarded from analysis either due to being in the shadow of the beamstop or their abnormal behaviors. The white ``+" indicates the beam center. Displayed in Fig.~\ref{fig:ave_scattering} (b) is the radial average of the averaged scattering. The decreasing intensity with increasing $Q$ values results from the form factor of the gold nano-particles. Fitting the radial average to the sphere form factor with Gaussian polydispersity yields the particle size in good agreement with the specifications provided by the vendor (Nanopartz). Limited by the detector coverage, we cannot obtain an analysis region with uniform intensity. In our analysis, we chose the annulus centered at $Q = 0.0055$~\AA$^{-1}$ as highlighted in white in Fig.~\ref{fig:ave_scattering}(a) (Suppl.) as our ROI. It is worth noting here that the count rate in the ROI is on average 0.01 photons/pixel. As a result, we cannot directly observe speckles in the single-shot scattering, and contrast analysis relies on photon statistics, i.e., measuring probabilities  of  multiple  photons  per  pixel events~\cite{sun2020accurate}.
Below we present the steps to correct the systematic statistical error of the contrast evaluation. 
First, we provide a derivation for correcting the additional contrast contribution from the form factor.
The total scattering intensity $I$ can be factorized into the form factor part $I_f$ that varies as a function of $Q$ and speckles $I_S$ that modulates the intensity  
$$
I = I_S  I_f
$$
As the two are independent, we have
$$
\mathrm{var}(I) =   \mathrm{var}(I_S) \mathrm{var}(I_f) + \bar{I_S}^2 \mathrm{var}(I_f) + \bar{I_f}^2 \mathrm{var}(I_S),
$$
and after dividing the equation by $\bar{I_S}^2 \bar{I_f}^2$, we arrive at
\begin{equation}
   \beta(I_S) = \frac{\beta(I)-\beta(I_f)}{1+\beta(I_f)}
   \label{eq:correction1}
\end{equation}
with 
$$
\beta(I) = \frac{\mathrm{var}(I)}{\bar{I}^2}, \,\,\,
\beta(I_S) = \frac{\mathrm{var}(I_S)}{\bar{I_S}^2}, \,\,\,
\beta(I_f) = \frac{\mathrm{var}(I_f)}{\bar{I_f}^2}.
$$
Here $\beta(I)$ is the contrast extracted directly from the ROI we pick and $\beta(I_S)$ is the contrast arising solely from speckles that links directly to Eq.~1 of the main paper. From the intensity distribution in the averaged scattering, the contrast contribution from the non-uniform intensity in the ROI was estimated to be $\beta(I_f)\approx 5\%$.

\subsection{Correction for low count rate photon assignment algorithms}
\begin{figure}[h!]
    \centering
    \includegraphics[width=0.5\linewidth]{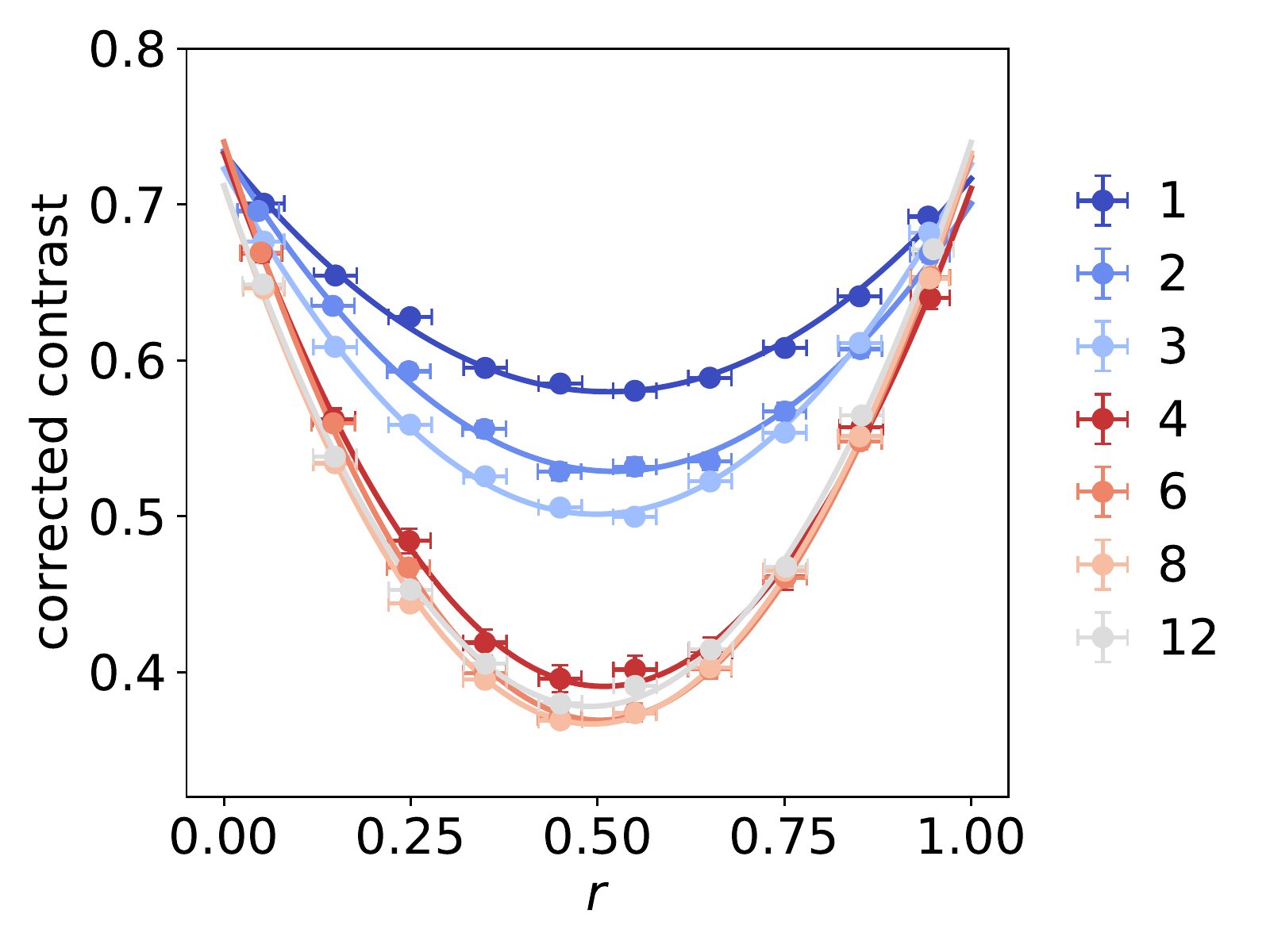}
    \caption{(Supplemental) Corrected contrasts for different jet flow rates for the ring ROI centering at $Q\approx0.0055$~\AA$^{-1}$ as a function of the first pulse intensity fraction $r$. The curves are the fitting results using Eq.~1 of the main paper.}
    \label{fig:contrast_degredation}
\end{figure}
We created the scattering sum by adding two random frames from the scattering data under the condition of the jet flow rate 1~mL/min. The chosen frames also satisfy $r<0.1$, and this is to ensure that the frames used have the same contrast level. In total 14000 of summed-scattering images are created and contrast is analyzed in the same ring as in Fig.~\ref{fig:contrast_correction} (Suppl.) using the greedy guess (GG) photon positioning algorithm~\cite{sun2020accurate}. The fitting to Eq.~1 of the main paper ($ |f(\bm{Q},\bm{v},\tau)|^2 = 1$) yields a contrast $\beta_e^s(r^s = 0.5) \approx 0.26\pm 0.01$. Here the superscript $s$ denotes the contrast from the artificial scattering sum and the subscript $e$ denotes the extracted contrast using GG. Based on the discussions in ~\cite{sun2020accurate}, the biased extracted contrast can be corrected using a linear model:
$$
\beta_e(I) = \alpha \beta_a (I) + \alpha -1,
$$
here $a$ in the subscript denotes the absolute contrast of the scattering intensity $I$. In order to extract the contrast of the speckles decorating the scattering, we need to use Eq.~\ref{eq:correction1}, and the formulation relating the actual contrast $\beta_a(I_S)$ and the extracted contrast $\beta_e(I)$ is
$$
\beta_e(I) = \alpha (1+\beta(I_f)) \beta_a (I_S) +  \alpha (1+\beta(I_f)) -1.
$$
From this we can define the effective correction factor $\alpha' = \alpha (1+\beta(I_f))$. And the correction for extracting the absolute contrast of speckles becomes
\begin{equation}
 \beta_a(I_S) = \frac{\beta_e(I) +1 -\alpha'}{\alpha'}
 \label{eq:correction2}
\end{equation}
As the recorded frames are at least 8.3 millisecond apart, the two random scatterings chosen are scattered from two completely different parts of the sample due to the jet flow, we have $\beta_a^s(r^s = 0.5) = 0.5\beta_a^s(r^s = 0) = 0.5 \beta_a^s(r = 1)$, this gives
$$
\frac{\beta_e^s(r^s = 0(1))+1-\alpha'}{\beta_e^s(r^s = 0.5)+1-\alpha'} = 2,
$$
yielding the effective correction coefficient $\alpha' \approx 0.94$ and the corrected contrast is then 
$\beta_a^s(r^s = 0)=\beta_a^s(r^s = 1)\approx 0.67$. 
Using this correction coefficient, we are able to correct the contrasts extracted at each flow rate for the ring ROI as a function of $r$, and this is displayed in Fig.~\ref{fig:contrast_degredation} (Suppl.). From this, we are able to extract $\beta_1$, $\beta_2$ as well as $|f|^2$, which are used to plot Fig.~2 of the main paper.
\section{III. Discussion of sample thickness}
Here we discuss the approximations in the theoretical derivation of the intermediate scattering function, related to the sample thickness of the jet. First, as mentioned, $\sigma_x\approx 1.8~\mu$m is much smaller than the jet diameter $d$, and we thus treated the sample to be uniformly thick in the integration. Moreover, we approximated that the velocity distribution of the particles flowing with the jet was only determined by $z$. 

The electric field also has a dependence on $z$ with
$$
E(\bm{r},\tau) = E_0 \exp[-\frac{x^2}{2\sigma_x^2}] \exp[-\frac{(y-\bar{v}\tau)^2}{2\sigma_y^2}]\exp[-\frac{z}{2z_0}],
$$
with $z_0$ being the attenuation length.  
Under the uniform flow condition, this factor gets cancelled as it exists in both the numerator and denominator for the calculation of intermediate scattering function. 
Under the condition where the jet has a velocity gradient, a more exact form of the intermediate scattering function should then be
\begin{equation*}
\begin{split}
    |f(\bm{Q},v, \tau)| & \approx  \frac{1}{\zeta_0(1-\exp[-2/\zeta_0])}\\ |\int_{-1}^{1}  &\exp[-(\zeta+1)/\zeta_0] \exp[i\frac{\tau}{\tau_2}2(1-\gamma)\zeta^2] \exp\{-\frac{\tau^2}{\tau_1^2}[2(1-\gamma)(1-\zeta^2)+\gamma]^2\} ~d\zeta|.  
\end{split}
\end{equation*}
Here $\zeta_0 = 2z_0/d$. For 8.2~keV photon energy, the attenuation length is 2.7~$\mu$m and 1.07~mm for gold and water respectively. Considering the volume percentage of the gold particle $\Phi = 0.026\%$, $z_0 \approx 41.3$~$\mu$m. 
Comparing numerically, Eq.~3 of the main paper gives less than 0.01 error bar of $|f|^2$ for the speed and $\gamma$ values in our measurement setup and thus we also ignored this factor.
\section{IV. Speckle pattern simulation}
The volume simulated spans across $20 \times 20 \times 92~\mu m^3$ along $\hat{\bm{x}},\hat{\bm{y}},\hat{\bm{z}}$ and is sampled with grid size of $10 \times 10 \times 100$~nm ($2000 \times 2000 \times 920$ grids). The grid along $\hat{\bm{z}}$ is larger as we are calculating small angle scattering ($Q \sim 0.01 \mathrm{\AA}^{-1}$, or scattering angle $2\theta \sim 0.1^{\circ}$) and thus is not sensitive to $\bm{\hat{z}}$. We randomly initialized the positions of 100000 particles and added the particles to the simulated volume. The construction of individual particles started with with $15\times15\times 1$ grids. The initial values of the grids were 0, and set to to 1 if a grid has distance less than 50~nm to their center. Due to the finite grid size, we subsequently added random rotations around $z$ to each particles to better mimic the spherical shape. Their velocities were along $\hat{y}$, and were calculated based on their $x$ and $z$ location in the volume assuming a parabolic profile with an average speed $\bar{v} = 0.5~$m/s. Note here that the coarse grid along $z$ didn't prevent us saving a finer position of particle $z$ position for the velocity calculation. The input beam has a Gaussian profile with a beam FWHM size of 3 $\mu m$ (300 grids) along $\hat{x},\hat{y}$, aligned to the center of the simulation volume. First we calculated the scattering from the sample at its initial state. After updating the positions of the particles in the scattering volume based on their velocities a time separation of $\tau = 49~$ns later, we calculated the intensity of the scattering again. We added the two scatterings together to create the speckle sum displayed in the manuscript. 

\bibliography{supplemental_reference}